\documentclass[Afour,sageh,times]{sagej}

\usepackage{moreverb,url}

\usepackage[colorlinks,bookmarksopen,bookmarksnumbered,citecolor=black,urlcolor=black]{hyperref}

\newcommand\BibTeX{{\rmfamily B\kern-.05em \textsc{i\kern-.025em b}\kern-.08em
T\kern-.1667em\lower.7ex\hbox{E}\kern-.125emX}}

\usepackage{balance}       % to better equalize the last page
\usepackage{graphics}      % for EPS, load graphicx instead 
\usepackage[T1]{fontenc}   % for umlauts and other diaeresis
\usepackage{txfonts}
\usepackage{mathptmx}
\usepackage{tabularx}
\usepackage{color}
\usepackage{colortbl}
\definecolor{lightgray}{gray}{0.9}
\usepackage{array}
\newcolumntype{L}[1]{>{\raggedright\let\newline\\\arraybackslash\hspace{0pt}}m{#1}}
\newcolumntype{C}[1]{>{\centering\let\newline\\\arraybackslash\hspace{0pt}}m{#1}}
\newcolumntype{R}[1]{>{\raggedleft\let\newline\\\arraybackslash\hspace{0pt}}m{#1}}
\newcolumntype{B}{@{\extracolsep{0.5cm}}c@{\extracolsep{0pt}}}%

\usepackage{booktabs}
\usepackage{textcomp}

\usepackage{multirow}
\usepackage{microtype}        % Improved Tracking and Kerning
\usepackage{ccicons}          % Cite your images correctly!
\usepackage{lscape}
\usepackage{graphicx}

\usepackage[normalem]{ulem}

\newcommand\clearrow{\global\let\rowmac\relax}
\clearrow

\definecolor{linkColor}{RGB}{6,125,233}

\def\volumeyear{2016}

\begin{document}
\newcommand{\subsubsubsection}[1]{\textbf{#1.}}

%\runninghead{Tawfiq Ammari}

\title{Normalized Surveillance in the Datafied Car:
How Autonomous Vehicle Users Rationalize Privacy Trade-offs}

% some titles (by chatgpt)
%    "Visual Rhetorics of Critical Race Theory: A Discourse Analysis of Facebook Memes"
%    "Political Epistemology of CRT Memes: Redefining Racism and Antiracism on Facebook"
%    "Decoding CRT Memes: Visual Culture and Political Mobilization in the Social Media Age"
%    "Beyond Post-Truth: Facebook Memes and the Political Epistemology of Critical Race Theory"
%    "Meaning-Making and Misinformation: A Study of Critical Race Theory Memes on Facebook"
\author{Yehuda Perry \affilnum{1}, Tawfiq Ammari\affilnum{1}}
%\author{Anonymous Authors}
\affiliation{Anonymous}
 \affiliation{\affilnum{1}Rutgers University, USA\\}

% \corrauth{Alistair Smith, Sunrise Setting Ltd
% Brixham Laboratory,
% Freshwater Quarry,
% Brixham, Devon,
% TQ5~8BA, UK.}

\email{tawfiq.ammari@rutgers.edu}

\begin{abstract}
Autonomous vehicles (AVs) are increasingly characterized by pervasive datafication and continuous surveillance through sensor arrays including in-cabin cameras, LIDAR systems, and GPS tracking. Drawing on 16 semi-structured interviews with AV drivers analyzed using constructivist grounded theory, this study examines how users make sense of vehicular surveillance within the broader context of everyday datafication. Our findings reveal that drivers demonstrate few AV-specific privacy concerns, instead normalizing vehicular monitoring through comparisons with established digital platforms. We theorize this privacy indifference by situating AV surveillance within the `surveillance ecology' of contemporary platform environments, arguing that the datafied car functions as a mobile extension of what scholars term the `leaky home' -- private spaces rendered permeable through connected technologies that continuously transmit behavioral data. The study contributes to scholarship on surveillance beliefs, datafication, and platform governance by demonstrating how users who have accepted comprehensive smartphone and smart home monitoring encounter AV datafication as simply another node in normalized data extraction. We further highlight how geographic restrictions on data access -- currently limiting driver log access to California residents -- create asymmetries that impede informed privacy deliberation, exemplifying `tertiary digital divides' in datafied environments. Finally, we examine how contemporary machine learning's reliance on data-intensive approaches creates structural pressure for comprehensive surveillance that transcends individual manufacturer choices. We propose governance interventions to democratize social learning about AV surveillance, including universal data access rights, binding transparency requirements, and data minimization standards that prevent race-to-the-bottom dynamics in automotive datafication.
\end{abstract}


\maketitle

\section{Introduction}

The automobile is being transformed from a mechanical object into a data-generating platform \citep{hind2024automotive}. Contemporary autonomous vehicles (AVs) collect vast data through pervasive sensors: cameras monitor driver attention, LIDAR maps environments, and GPS tracks movement patterns \citep{hind_dashboard_2021}. These streams flow continuously to manufacturers, fueling machine learning algorithms \citep{stilgoe2018machine}. Yet unlike earlier automotive safety technologies, today's AVs participate in what \citet{zuboff_2015} terms 'surveillance capitalism' -- economic systems premised on extracting and commodifying behavioral data.

This datafication raises pressing questions about privacy. How do AV drivers understand pervasive vehicular monitoring? Drawing on 16 semi-structured interviews with AV users, we examine how drivers make sense of surveillance practices. Our analysis reveals a striking paradox: despite extensive monitoring systems, participants expressed minimal AV-specific privacy concerns, instead rationalizing data collection through comparisons with established digital platforms.

This pattern echoes what \citet{draper2019corporate} theorize as `digital resignation' -- a condition when people desire to control their digital information but feel unable to do so. Unlike the `privacy paradox' that frames user behavior as irrational, digital resignation conceptualizes acquiescence as rational response to corporate surveillance that systematically cultivates helplessness.

These findings demand attention to normalization processes in platform-mediated environments. AVs exemplify what \citet{hind_dashboard_2021} calls the `datafied driving experience' -- reconfiguring automobility wherein data extraction becomes naturalized. Building on scholarship examining surveillance beliefs \citep{segijn_my_2025}, we demonstrate how platform rhetoric persuades users to accept monitoring they might otherwise resist. \citet{rosso2020chilling} document surveillance awareness producing measurable behavioral changes, yet our participants exhibited resignation rather than resistance, potentially reflecting successful normalization of automotive surveillance. This has governance implications: if social learning about AV surveillance is privatized \citep{stilgoe2018machine}, manufacturers gain unilateral power to define privacy boundaries.

Our analysis engages with scholarship on agency in datafied societies \citep{hepp2024agency}, examining specific practices through which AV users exercise -- or fail to exercise -- agency over vehicular data.

\subsection{Research Question}

This study addresses:

\begin{quote}
\textbf{RQ:} How do autonomous vehicle drivers conceptualize privacy concerns in the context of continuous data collection, and how do they rationalize their participation in vehicular surveillance?
\end{quote}

By examining user sense-making around AV data practices, we contribute to scholarship on datafication, platform governance, and algorithmic technologies. Our findings suggest privacy indifference represents not user ignorance but successful platform persuasion --- what \citet{lutz2020data} describe as `privacy cynicism,' an attitude of uncertainty, powerlessness, and mistrust rendering privacy protection subjectively futile.

\section{Theoretical Framework}

\subsection{Datafication and the Automotive Platform}

Contemporary vehicles function as platforms -- infrastructures mediating relationships between actors while extracting value from user interactions \citep{hind2024automotive}. This platformization transforms automobiles into networked ecosystem nodes, enabling novel data capture \citep{hind_dashboard_2021}. Dashboard redesigns exemplify these processes: converging previously separate interfaces into unified digital cockpits that continuously generate behavioral data \citep{hind_dashboard_2021}.

Datafication -- `the transformation of human life into data through quantification' \citep{mejias2019datafication} -- operates at multiple registers in AVs. External sensors transform environments into navigational data; internal monitoring converts driver behaviors into training data \citep{hind_dashboard_2021, stilgoe2018machine}. \citet{hind_dashboard_2021} identifies two key design techniques: (1) representational transparency, depicting interfaces as seamlessly responding to `natural' behaviors, and (2) customizable control, offering superficial agency while deepening extraction.

\citet{barassi2020datafied} identifies three interconnected temporalities characterizing surveillance capitalism: \textit{immediacy} (constant connectivity pressure); \textit{archival time} (transforming present into permanent records); and \textit{predictive time} (using historical data to anticipate futures). These prove relevant to AVs, where immediate driving data is archived indefinitely for algorithmic improvement and user behavior prediction.

\subsection{Surveillance Capitalism and Digital Resignation}

\citet{zuboff_2015} argues surveillance capitalism transforms human experience into behavioral data for prediction markets, manufacturing consent through what \citet{agre1995soul} calls `empowerment and measurement regimes' -- systems granting limited agency while enabling comprehensive monitoring. AVs exemplify this: drivers receive autopilot capabilities in exchange for in-cabin cameras and data sharing.

Scholarship on surveillance beliefs \citep{segijn_my_2025} shows users often hold inaccurate mental models about data collection. These beliefs shape behaviors: surveillance consciousness can produce `chilling effects' \citep{penney2021understanding}. \citet{rosso2020chilling} provide empirical evidence, documenting measurable internet search behavior changes following Snowden revelations, distinguishing between subjective privacy concerns and objective behavioral changes.

\citet{draper2019corporate} theorize `digital resignation' as rational response to corporate surveillance that systematically cultivates helplessness -- reframing the `privacy paradox' as strategic corporate cultivation. They identify key strategies: obfuscation through complex policies, diversionary offers of superficial control, and `seductive surveillance' emphasizing conveniences. \citet{lutz2020data} introduce `privacy cynicism,' finding greater privacy awareness paradoxically increases powerlessness. \citet{mcguigan2023after} extend this to corporate behavior, identifying `cynical resignation' wherein companies perform privacy compliance while preserving extraction through: `sanitizing surveillance' via computational obfuscation, `party-hopping' converting third-party to first-party data, and `sabotage' disadvantaging competitors.

\subsection{Machine Learning and Privatized Social Learning}

\citet{stilgoe2018machine} theorizes AV development as dual machine learning and social learning processes. While algorithms learn to navigate, society learns about these technologies through discourse, regulation, and user experiences. Crucially, much social learning has been privatized: manufacturers control data flows, define learning objectives, and resist governance constraints. This creates governance challenges, as machine learning opacity -- both technical and strategic -- limits accountability \citep{stilgoe2018machine}.

\citet{stilgoe2018machine} advocates governance prioritizing social learning through data sharing requirements and democratic algorithmic engagement, resisting techno-solutionist framings treating privacy as engineering rather than political problems. Contemporary machine learning's transition from `data-poor clever algorithms to data-rich simple ones' places data extraction at AI development's center, creating structural incentives for comprehensive surveillance \citep{stilgoe2018machine}.

\subsection{Everyday Datafication and the Surveillance Ecology}

\citet{sophus2020proxy} develop methodological approaches for examining mobile apps' `surveillance ecology' -- infrastructural components conditioning user data disclosure. Their analysis reveals privacy implications emerge not from individual apps but from cumulative extraction across platforms through interconnected technical systems: SDKs embed tracking, advertising networks coordinate data flows, and APIs enable behavioral profiles traveling between corporate actors.

AVs extend this ecology into automotive contexts with similar permissions, data flows, and third-party relationships. Just as smartphone apps request location access while obscuring how data travels to advertising networks, AV systems present driver monitoring as safety functionality while embedding commercial extraction capabilities. \citet{barassi2020datafied} demonstrates how these ecologies produce `techno-dependency' -- conditions where constant connectivity becomes functionally mandatory. In her family technology ethnography, daily institutions increasingly require digitized engagement, leaving `no choice but to constantly be connected.' The AV context intensifies this: to access advanced safety features, drivers must accept comprehensive monitoring.

\citet{soilen2025leaky} propose the `leaky home' as conceptual figure for understanding how automated domestic spaces transmit data through connected devices. The datafied car extends this leakiness into mobile contexts -- a private space continuously transmitting behavioral data through public environments. Unlike the leaky home (spatially fixed), the AV creates a mobile surveillance node generating location data wherever it travels while recording cabin activities.

\citet{urquhart2022policing} conceptualize smart home devices as `invisible witnesses' that become domesticated into daily routines while accumulating trace evidence data. AVs function similarly, recording driving behavior, conversations, locations, and mundane vehicular life rhythms.

\citet{cheong2022smart} examine how users navigate datafication in smart campus environments, identifying `tertiary digital divides' -- asymmetries in users' ability to understand and respond to datafication compounding existing inequalities. Users who cannot access or interpret their own data remain unable to develop informed mental models, limiting capacity for meaningful privacy deliberation.

\citet{hepp2024agency} situate these experiences within broader debates about agency in datafied societies, emphasizing that agency questions must be examined in relation to particular sociotechnical contexts. Agency `is always defined by its social localization and situational realization,' including language, emotion, and power relations.

\subsection{Contextual Privacy and AI Trade-offs}

\citet{solove2010understanding} argues privacy must be understood contextually -- in relation to specific informational flows, social relationships, and power dynamics. Building on \citet{nissenbaum2004privacy}, we analyze AV surveillance through contextual integrity: privacy violations occur when information flows violate appropriate informational norms for given contexts. AVs complicate these norms because they blur boundaries between public and private spaces.

\citet{tang2025between} examine how users navigate risk-benefit trade-offs in AI adoption through `self-negotiation,' weighing perceived benefits against concerns through either promotion-focused orientations (emphasizing benefits while downplaying risks) or prevention-focused orientations (prioritizing privacy protection). Platform design exploits this dynamic by emphasizing empowerment while obscuring measurement -- creating conditions where promotion-focused self-negotiation appears rational even as it serves corporate interests.

\section{Methods}

Following IRB approval, 16 regular AV drivers (US residents, 18+, using AVs 3+ times weekly for six months) were recruited via social media and Volunteer Science, a web-based platform with over 50,000 registered volunteers \citep{radford2016volunteer}. The sample included Tesla Model 3/Y/S (n=12), Polestar 2 (n=2), Chevrolet Bolt (n=1), and Mitsubishi Outlander (n=1) drivers (Table \ref{tab:demo}).

\begin{table*}[h!]
\centering
\caption{Participant demographics}
\label{tab:demo}
\small
\begin{tabular}{llllll}
\toprule
ID & Gender & State & Education & Vehicle & AI System \\
\midrule
P1 & F & WA & Graduate & Chevrolet Bolt & Autopilot \\
P2 & F & WA & Graduate & Tesla Model Y & FSD Beta \\
P3 & M & CA & Bachelor's & Tesla Model 3 & FSD Beta \\
P4 & M & NJ & N/A & Tesla Model 3 & FSD Beta \\
P5 & M & GA & High School & Tesla Model 3 & FSD Beta \\
P6 & M & TX & Bachelor's & Polestar 2 & Pilot Assist \\
P7 & M & NJ & Ph.D. & Polestar 2 & Pilot Assist \\
P8 & M & NJ & Graduate & Tesla Model 3 & FSD Beta \\
P9 & M & MD & Bachelor's & Tesla Model 3 & FSD Beta \\
P10 & M & TN & Graduate & Tesla Model Y & FSD Beta \\
P11 & F & CA & Ph.D. & Tesla Model S/X & FSD \\
P12 & M & VA & Graduate & Tesla Model 3 & FSD Beta \\
P13 & M & WA & High School & Tesla Model Y & FSD Beta \\
P14 & M & TX & High School & Mitsubishi & M-Pilot Assist \\
P15 & M & NJ & Graduate & Tesla Model 3 & FSD \\
P16 & M & NC & Bachelor's & Tesla Model 3 & FSD Beta \\
\bottomrule
\end{tabular}
\end{table*}

Semi-structured interviews (45-90 minutes) explored AV use, privacy configurations, and data collection awareness. The complete protocol appears in the supplementary appendix. Interviews were transcribed using Otter AI, reviewed for accuracy, and imported into NVivo for analysis.

Guided by \citet{charmaz2006constructing}, we adopted constructivist grounded theory methodology, moving from open coding to focused coding. As noted by \citet{clarke2003situational}, this framework suits examining emergent sense-making in sociotechnical environments, enabling categories to arise inductively from participant narratives \citep{saldana2011fundamentals}. Both authors engaged in multiple focused coding rounds via constant comparison \citep{charmaz2006constructing}. Following \citet{braun2006using}, we approached themes as active constructions resulting from researcher engagement with participant narratives, aligning with critical data studies \citep{barad2007meeting, kennedy2016post}.

Final analysis identified four primary privacy-related themes (Table \ref{tab:themes}).

\begin{table}[h!]
\centering
\caption{Privacy themes, number of interviews, and frequency}
\label{tab:themes}
\begin{tabular}{lcc}
\hline
Theme & Interviews & Instances \\
\hline
Justified data collection & 11 & 16\\
Technological necessity & 10 & 14 \\
Trust in manufacturers & 14 & 17\\
Platform comparison & 15 & 20\\
\hline
\end{tabular}
\end{table}

\section{Findings}

\subsection{Minimal AV-Specific Privacy Concerns}

Only three participants expressed specific privacy concerns about their AVs, and even these remained abstract rather than concrete. P7, an academic studying technology's social effects, noted: `I mean, you never know how the data that is being collected is going to be used...and the possibility that it can be used against your own interests is always there.' This anxiety remained generalized -- P7 could not identify specific AV data practices that troubled them. This pattern mirrors what \citet{lutz2020data} describe as privacy cynicism: awareness of risk coexisting with inability to articulate specific threats or effective protective actions.

P11, a privacy researcher, was the sole participant to demonstrate detailed privacy consciousness. She expressed concerns about Tesla's Sentry Mode, which 'stores [images of] anyone who gets close to the vehicle in case the car gets hit when it's parked. And I really don't think people understand that...they are being recorded if they are near the car.' Notably, P11's concern centered on surveillance of \textit{others} (pedestrians) rather than herself -- a finding consistent with scholarship on third-party privacy in smart devices \citep{emami2019exploring}. A critical nuance in user resignation to surveillance was the conflation of "outward" sensors (mapping the world) with "inward" monitoring (watching the driver). P8 noted the significant technical shift when Tesla removed radar sensors in favor of "Tesla Vision," a camera-only approach that necessitates higher fidelity visual processing. While users like P7 expressed high trust in the engineering required for external safety—believing the car needs to see the road—this trust often acted as a Trojan horse for internal surveillance. Participants frequently justified cabin cameras (inward surveillance) using the logic of road sensors (outward surveillance), failing to distinguish between the machine's need to navigate the environment and its design to monitor their personal attentiveness. This blurring of boundaries allows manufacturers to introduce invasive biometric monitoring under the unassailable banner of "safety engineering."

Participants attempted to retrieve their vehicle telematics via Tesla's "Data Privacy" portal, a web-based feature allowing owners to request a "copy of your data" including vehicle usage statistics, service history, and safety event recordings \citep{tesla_privacy}. The standard retrieval process involves navigating to the "Data \& Privacy" menu in the user account, selecting "Obtain a Copy of Your Data," and awaiting a compilation period of up to 30 days \citep{mozilla_2023}.

P11 was one of two California residents in our sample and, consequently, the only participant with successful access to these logs. When we asked her to review the returned data, she expressed surprise at the detail: ``I did not expect to see this level of detail.'' Records included door open/close timestamps, tire pressure readings, and second-by-second velocity data. Despite acknowledging the granularity, P11 struggled to articulate specific harms, echoing earlier research showing gaps between privacy awareness and risk perception \citep{gerber2018explaining}. Critically, the other participants in our sample could not access their own data logs -- a geographic restriction that prevented them from developing similar awareness of surveillance granularity. This asymmetry exemplifies what \citet{cheong2022smart} describe as `tertiary digital divides,' where unequal access to data visibility compounds existing inequalities in technological literacy. The inability to access one's own data forecloses the possibility of developing accurate surveillance beliefs -- a precondition for meaningful agency in datafied contexts \citep{hepp2024agency}.

P12 captured the prevailing sentiment: `You know, I have not really worried about [the privacy of my data in the AV]. Maybe I should be but I'm not.' The remaining thirteen participants shared this indifference to AV-specific privacy concerns. Instead, they rationalized data collection through three primary frames: security/trust in manufacturers, technological necessity, and platform equivalence. These rationalizations align with \citet{draper2019corporate}'s theorization of digital resignation as a cultivated phenomenon: users develop explanatory frameworks that justify acquiescence rather than challenge surveillance practices.

\subsection{Rationalizing Surveillance: Three Justificatory Frames}

\subsubsection{Trust in Platform Security}

Many participants expressed confidence in manufacturers' data security practices. P16 read Tesla's privacy agreement and concluded: `everything sounds like Tesla is the [only entity that's] collecting the data. And it doesn't say anywhere that they're sharing it with third parties.' P2 similarly stated: `I really feel like the propriety that Tesla offers to its customers is pretty ironclad. I don't feel like they share the information.' This trust extended to comparisons with established technology companies: `Apple kind of has the same thing. It's really difficult to crack Apple.'

These responses demonstrate what \citet{hind_dashboard_2021} calls `representational transparency' -- design strategies that depict platforms as seamlessly protecting user interests. By positioning themselves as privacy-conscious technology companies rather than traditional automakers, Tesla and others leverage trust cultivated in the consumer electronics sector. Notably, participants distinguished between \textit{data collection} (acceptable) and \textit{data sharing} (concerning), despite both practices serving corporate interests in surveillance capitalism \citep{zuboff_2015}. This distinction reflects what \citet{mcguigan2023after} identify as successful `sanitizing surveillance': corporate strategies that reframe comprehensive monitoring as protective rather than extractive. Participants trusted that their data remained `safe' within Tesla's ecosystem while remaining unable to verify this claim or understand how collected data might be used.

P1 ignored the vehicle's Wi-Fi hotspot features entirely, assuming they lacked the security of her home network. Despite this variance in active usage, five participants believed third parties were already accessing AV data, though they struggled to identify who or for what purposes. This mirrors \citet{segijn_my_2025}'s findings on electronic eavesdropping beliefs: users sense pervasive surveillance without accurate understanding of specific data flows. The difference lies in affect -- whereas e-eavesdropping beliefs generate anxiety, AV users accepted third-party access as inevitable. This affective shift from anxiety to acceptance characterizes the transition from privacy concern to digital resignation \citep{draper2019corporate}. Where surveillance awareness might once have produced chilling effects \citep{rosso2020chilling}, our participants demonstrated normalized acceptance -- suggesting that automotive contexts may be particularly conducive to resignation cultivation.

\subsubsection{Technological Necessity: Training Better Algorithms}

Ten participants viewed data sharing as essential for algorithmic improvement. P3 stated: 'if they're collecting data, I assume they're collecting it for reliability and reliability related issues.' This frame naturalizes datafication as a technical requirement rather than a business model choice \citep{zuboff_2015}. As \citet{barassi2020datafied} observes, the temporalities of surveillance capitalism create conditions where constant data generation appears necessary rather than optional -- an `onlife' existence where digital and physical realms merge inseparably.

Critically, several participants noted they had \textit{no choice} in data sharing. Full Self-Driving (FSD) Beta requires consenting to in-cabin camera monitoring -- a non-negotiable condition. P2 explained: `The in-car cabin camera is not optional if you're going to [use] full self-driving beta...you have to give the cameras access. So that's non-negotiable.' P11 confirmed: `I feel not super happy about this' but she still accepted the terms to access the latest AI capabilities. Participants often framed their acceptance of surveillance not as a choice, but as a submission to a technological monopoly where no viable "opt-out" exists for those seeking advanced safety. Robert Peterson vividly described this lack of agency, noting that once the technology matures, it effectively renders traditional driving obsolete, yet the cost is total data submission. P7 argued, "If nobody else embraces this technology... The barrier to [the creation of an automated self-driving car] is insurmountable," suggesting a future where non-surveilled transport is not just inconvenient, but structurally impossible. This sentiment reinforces the theme of "forced datafication," where the utility and safety benefits of AVs function as a lock-in mechanism. Users perceive the trade-off as a non-negotiable contract: to access the physical safety of the vehicle, one must surrender to the digital surveillance of the manufacturer, creating a "no exit" scenario for modern mobility. This experience of compulsion echoes \citet{draper2019corporate}'s analysis of how corporate practices cultivate resignation by constructing surveillance as inevitable -- presenting monitoring not as a corporate choice but as a technological necessity that users must accept to participate in modern life.

This exemplifies \citet{agre1995soul}'s `empowerment and measurement regime': users receive enhanced capabilities (better autopilot) in exchange for comprehensive monitoring. Participants framed this as a fair trade-off. P3 suggested: `I don't even think about it because all the benefits outweigh any of the negatives...by far.' These responses align with \citet{tang2025between}'s findings on risk-benefit trade-offs in AI usage: users engage in `self-negotiation,' often adopting promotion-focused orientations that emphasize benefits while downplaying privacy risks. \citet{lutz2020data} would characterize this as privacy cynicism: participants recognize surveillance exists yet conclude that protective behavior is futile, leading them to emphasize benefits as a coping mechanism rather than a reasoned calculation.

However, as \citet{stilgoe2018machine} argues, framing data collection as technically necessary obscures political choices. Manufacturers could potentially train algorithms using simulation, anonymized fleet data, or opt-in testing rather than mandatory monitoring. By presenting comprehensive surveillance as the \textit{only} path to better AI, they privatize social learning and foreclose democratic deliberation about acceptable data practices. The contemporary machine learning paradigm's reliance on `data-rich simple algorithms' \citep{stilgoe2018machine} creates structural pressure for maximum data extraction regardless of whether specific data points serve legitimate safety purposes.

Beyond passive observation, the datafied car functions as a disciplining agent—a "Panopticon on wheels"—where users actively modify their behavior to appease the algorithm. Participants described a transactional relationship with the vehicle, where "good" behavior is currency. Mitchell, a police officer, detailed his meticulous tracking of FSD Beta disengagements, treating them not just as software errors but as strikes against his record as a "worthy" beta tester. Participants in this study frequently referred to `FSD Beta' (Full Self-Driving Beta), a developmental tier of Tesla's automation software that extends driver assistance from highways into complex city environments (e.g., stopping at traffic lights, navigating roundabouts, and making unprotected left turns). Unlike standard cruise control, FSD Beta is explicitly unfinished software released to a select group of public users for real-world validation. To access this feature, drivers often had to opt-in to invasive pre-monitoring, achieving a high `Safety Score' based on metrics like braking intensity and turning aggression \citep{TeslaSafetyScore}. Once admitted, these users effectively transition from customers to unpaid quality assurance testers. They consent to having their vehicle continuously record and upload high-fidelity video and telemetry data to the manufacturer—particularly when they intervene to correct the car's errors. This arrangement turns the private vehicle into an active node in the manufacturer's neural network training loop, where the driver's primary role is to supervise the machine and generate data on 'edge cases' that the software cannot yet handle.

This reveals that surveillance in AVs is not merely extractive but performative; drivers self-discipline their movements, speed, and attention to maintain high "Safety Scores" or avoid insurance premiums. The car does not just watch the road; it trains the driver to conform to the machine's definition of safety. As described by Stilgoe (2018), such systems represent a form of 'social learning' where the technology is tested in the public domain rather than closed circuits. This system effectively disciplines drivers to conform to algorithmic standards of 'good behavior' in exchange for software access \citep{Quartz2021}. Once admitted, these users transition into what scholars describe as unpaid quality assurance testers; they consent to continuous data extraction to train the manufacturer's neural networks, intervening to correct errors in what acts as a feedback loop of 'shadow labor' for the corporation. This behavioral modification represents what \citet{rosso2020chilling} might recognize as a chilling effect -- though here the effect serves corporate rather than governmental interests. Users modify their driving behavior not because they fear state surveillance but because algorithmic evaluation shapes access to valued features.

\subsubsection{Platform Equivalence: `Google Already Knows Everything'}

Thirteen participants rationalized AV surveillance through comparisons with digital platforms they already use. P9 stated: `yeah, I don't care. I don't do anything that I wouldn't want in the Washington Post.' P6 expanded: `I'm really not concerned with privacy. We've been online since 1995. I bought things online. Credit cards, everything. So I know all my information is out there somewhere.'

This normalization reflects what \citet{hind_dashboard_2021} terms the `datafied driving experience' -- a transformation wherein vehicular data extraction becomes just another node in users' existing surveillance networks. P11 articulated this clearly: `the phone goes even more places with me, right? And it has a whole bunch of apps on it that are collecting data on me.' For these users, AVs simply extend familiar platform dynamics into automobility. This finding resonates with \citet{sophus2020proxy}'s analysis of the `surveillance ecology' of mobile platforms: users who have accepted comprehensive smartphone monitoring encounter AV datafication as simply another extension of normalized data extraction. Following \citet{barassi2020datafied}, we can understand this normalization as the product of surveillance capitalism's temporal logics: participants have internalized the `immediacy' and `archival time' of digital platforms, making AV monitoring feel continuous with existing datafied life rather than distinct from it.

Interestingly, some participants recognized AVs collect \textit{more} granular data than smartphones. P3 noted: 'they know where I shop. They know where I eat. They know a lot more than probably Google does...because Google is monitoring you by your phone app only...I can still stop location sharing on the phone, but I can't do it in the car.' Despite this awareness, P3 remained unconcerned, suggesting the sheer ubiquity of platform surveillance has dulled privacy sensibilities. \citet{soilen2025leaky} describe this phenomenon through the metaphor of the `leaky home' -- private spaces that become permeable through connected devices. The datafied car represents a mobile extension of this leakiness, transmitting intimate behavioral data as it moves through public and private spaces alike. P3's awareness combined with inaction exemplifies \citet{lutz2020data}'s privacy cynicism: recognition that AV surveillance exceeds other platforms yet resignation to its inevitability renders this knowledge behaviorally inconsequential.

These findings support \citet{segijn_my_2025}'s argument that surveillance beliefs are contextually shaped. Whereas their participants expressed anxiety about e-eavesdropping, our AV users demonstrated resignation. The difference may lie in visibility: conversations are perceived as private (making listening particularly intrusive), while driving occurs in public spaces where monitoring feels more legitimate. Alternatively, the difference may reflect successful platform rhetoric: by framing surveillance as beneficial (better AI) rather than exploitative (behavioral prediction markets), manufacturers cultivate acceptance \citep{zuboff_2015}. As \citet{mcguigan2023after} observe in their analysis of adtech's `pivot to privacy,' companies increasingly perform privacy concern while preserving extractive practices -- a `cynical resignation' that shapes user expectations about what privacy protection can realistically entail.

\section{Discussion}

\subsection{Normalized Surveillance and Platform Persuasion}

Our findings reveal AV users have thoroughly internalized surveillance capitalism's logics, actively rationalizing monitoring through frames serving manufacturer interests: trust in corporate security, faith in technological necessity, and normalization through platform comparison. Following \citet{draper2019corporate}, this internalization reflects successful corporate cultivation of digital resignation, wherein people desire privacy but feel unable to achieve it, developing justificatory frameworks rendering acquiescence sensible.

Dashboard design plays a crucial role \citep{hind_dashboard_2021}. By integrating monitoring within seamless interfaces emphasizing user control, manufacturers persuade drivers to accept surveillance they might otherwise reject. Voice navigation positions speech as unmediated while requiring specific linguistic protocols generating training data. Customizable displays offer superficial agency while deepening extraction. \citet{mcguigan2023after} identify analogous dynamics where companies develop `privacy-preserving' solutions maintaining extractive practices -- what they term `sanitizing surveillance.'

These strategies align with \citet{agre1995soul}'s empowerment and measurement regimes. Users receive genuine capabilities but access requires comprehensive monitoring, creating psychological pressure to consent. As \citet{stilgoe2018machine} observes, such arrangements privatize learning, allowing manufacturers to unilaterally define acceptable privacy trade-offs. \citet{barassi2020datafied} documents similar dynamics where parents experience `no choice' in engaging with datafied institutions, creating conditions of mandatory rather than optional disclosure. The AV context intensifies this: to access technologies marketed as safer, users must accept surveillance as prerequisite.

\subsection{From Privacy Paradox to Digital Resignation}

Our findings challenge the `privacy paradox' framework pathologizing user behavior as irrational. Following \citet{draper2019corporate}, participants rationally perceived privacy protection as impossible and developed justificatory frameworks accordingly. This aligns with \citet{lutz2020data}'s privacy cynicism: greater awareness produces resignation rather than resistance. Participants knew AVs collect more data than smartphones yet accepted it nonetheless. \citet{hepp2024agency} offer a framework avoiding both technological determinism and passivity assumptions: our participants exercised constrained agency under substantial asymmetry.

\subsection{Governance Implications}

Privatization of social learning poses governance challenges. Participants' indifference reflects platform persuasion and resignation rather than informed consent. \citet{rosso2020chilling} show surveillance awareness produces behavioral changes only when users perceive alternatives; our participants see no path to non-surveilled mobility.

Building on \citet{stilgoe2018machine}, we propose four interventions. First, mandate data transparency through visual diagrams and contextual notices. However, \citet{draper2019corporate} caution transparency alone may be insufficient without substantive collection limits.

Second, establish data minimization requirements. \citet{stilgoe2018machine} shows technical necessity claims serve economic interests; regulations should require demonstrating legitimate safety purposes with opt-in defaults. \citet{mcguigan2023after} warn without minimization, companies will develop nominally compliant solutions preserving extraction.

Third, ensure meaningful data access. Geographic restrictions exemplify what \citet{sophus2020proxy} identify as the `surveillance ecology' conditioning data disclosure invisibly. \citet{cheong2022smart} demonstrate such asymmetries create `tertiary digital divides' impeding informed mental models. Universal access, deletion rights, and retention limits are needed. \citet{hepp2024agency} suggest meaningful agency requires not just data access but institutional conditions enabling effective response.

Finally, address third-party access. \citet{zuboff_2015} shows surveillance capitalism depends on opaque data markets. \citet{mcguigan2023after} document `party-hopping' strategies circumventing restrictions; governance must anticipate such evasive maneuvers.

\subsection{Design Interventions}

Design interventions can make privacy risks salient. Drawing on \citep{monreale2014privacy,lane2014privacy}, we propose visual representations rather than lengthy policies. In-cabin screens could show real-time indicators when cameras or microphones activate. Earlier research demonstrates simple visual cues effectively raise privacy consciousness \citep{haley2005nobody}. However, such interventions must be assessed against \citet{rosso2020chilling}'s findings: if awareness produces behavioral modification serving corporate interests rather than meaningful protection, visibility may reinforce surveillance.

For external surveillance, pedestrian-facing displays could indicate Sentry Mode activation. \citet{urquhart2022policing} conceptualize smart devices as `invisible witnesses' that become domesticated into everyday life, naturalizing surveillance. Design interventions making AV monitoring visible could interrupt this domestication, creating opportunities for informed deliberation.

\citet{hind_dashboard_2021} suggests AV dashboards could display data flows themselves -- showing users what information moves to servers in real-time. \citet{barassi2020datafied} notes families who became aware of datafication's temporal dimensions began actively negotiating with surveillance systems, suggesting visibility can produce resistance under appropriate conditions.

Design interventions must be mandated through regulation. Manufacturers have no incentive to voluntarily reduce data extraction. As \citet{stilgoe2018machine} observes, contemporary machine learning has embraced a `brute force' approach depending on ever-expanding data collection. This creates structural pressure for comprehensive surveillance regardless of whether specific data points serve legitimate safety purposes. \citet{tang2025promise} demonstrate how users engage in `self-negotiation,' often downplaying privacy concerns for technological gratification -- a dynamic manufacturers exploit. Governance should establish minimum transparency thresholds, preventing race-to-the-bottom dynamics. \citet{mcguigan2023after} emphasize that without binding requirements, companies will develop `privacy-preserving' technologies maintaining extraction while performing compliance.

\subsection{Situating AV Surveillance within Everyday Datafication}

Our findings resonate with broader scholarship on privacy in connected environments. \citet{soilen2025leaky} propose the `leaky home' for understanding how automated domestic spaces transmit data. The datafied car represents a mobile extension -- a private space continuously transmitting behavioral data. Just as smart home devices become `invisible witnesses' \citep{urquhart2022policing}, AVs witness mundane driving aspects.

\citet{barassi2020datafied}'s analysis illuminates how this witnessing operates across time. AVs participate in all three identified temporalities: \textit{immediacy} (real-time monitoring requiring constant connectivity); \textit{archival time} (indefinite storage transforming present into permanent records); and \textit{predictive time} (algorithmic analysis using historical data to anticipate futures). This means AV surveillance is persistent across time -- creating longitudinal profiles following users from purchase through every driving moment.

This situating explains our participants' indifference. Users who have accepted the `leaky home' and smartphone surveillance ecology \citep{sophus2020proxy} encounter AV monitoring as another extension of normalized extraction. The platform comparison frame reflects genuine continuity between AV datafication and the broader digital ecosystem. As \citet{draper2019corporate} argue, this consistency reflects successful cultivation: corporations have established surveillance as baseline expectation across contexts.

Governance interventions must address not just AV-specific practices but the broader surveillance infrastructure normalizing comprehensive monitoring. \citet{mcguigan2023after} contend that without interventions into underlying business models, companies will continue developing technical solutions sidestepping privacy requirements. The `cynical resignation' they identify among corporations mirrors the digital resignation our participants express -- suggesting a troubling equilibrium where both users and companies accept surveillance as inevitable while performing concern about privacy.

\section{Limitations and Future Directions}

Our study examined 16 US-based AV users -- a limited sample that cannot represent diverse populations. Future research should analyze privacy attitudes across cultural contexts, particularly in jurisdictions with stronger privacy protections \citep{sheth_et_al_14}. \citet{lutz2020data}'s research with German users suggests privacy cynicism operates across national contexts, but variations in regulatory environment and cultural attitudes may produce distinct patterns.

Interview methods may underestimate privacy concerns. \citet{malkin2019privacy} showed users conceive more specific anxieties when presented with detailed scenarios. Future studies could employ vignettes depicting particular data uses to elicit more nuanced attitudes. \citet{rosso2020chilling} demonstrate the value of behavioral measures for documenting surveillance effects beyond self-reported attitudes. Future AV research could examine behavioral indicators like feature adoption rates or driving pattern modifications following privacy disclosures.

While datafication and surveillance capitalism provide powerful lenses, they may oversimplify complex user reasoning. Some participants may genuinely value data-driven improvements and make informed choices prioritizing functionality. \citet{hepp2024agency} emphasize that agency in datafied societies requires examination of specific sociotechnical contexts; future research should investigate how particular AV features, manufacturer practices, and user characteristics shape possibilities for meaningful privacy decision-making.

A significant limitation concerns data access asymmetries. Only our California-based participant could access her Tesla logs, precluding systematic analysis of how log review shapes privacy attitudes. Future research should examine whether data access influences user understanding \citep{cheong2022smart}. Additionally, our study preceded generative AI expansion into automotive contexts; future work should examine how large language model integration affects user privacy attitudes \citep{tang2025between}. \citet{draper2019corporate} argue that understanding resignation requires examining corporate practices that cultivate helplessness; longitudinal research tracking privacy attitudes alongside manufacturer communications could illuminate these dynamics.

Our study also preceded recent developments in AV regulation and public discourse. Major accidents, data breaches, and privacy scandals may shift user attitudes. \citet{mcguigan2023after} document how the advertising technology industry's `pivot to privacy' represents a new phase in surveillance capitalism; similar pivots in automotive contexts could reshape user expectations and manufacturer practices.

\section{Conclusion}

Drawing on 16 interviews with AV drivers, this study examined how users make sense of pervasive vehicular surveillance. We found drivers express minimal AV-specific privacy concerns, instead rationalizing data collection through comparisons with established digital platforms, illustrating surveillance capitalism's success in naturalizing monitoring as necessary for technological functionality.

Theoretically, our analysis contributes to scholarship on datafication \citep{hind_dashboard_2021}, surveillance beliefs \citep{segijn_my_2025}, and platform governance \citep{hind2024automotive, stilgoe2018machine}. We demonstrate how dashboard design persuades users to accept surveillance through empowerment and measurement regimes \citep{agre1995soul}, and how platform comparisons normalize data extraction. Privatization of social learning enables manufacturers to define privacy boundaries without democratic input. Building on \citet{draper2019corporate}'s theorization of digital resignation, we argue user acquiescence reflects rational response to corporate practices systematically cultivating helplessness, shifting analytical attention from individual decisions to structural conditions constraining those decisions.

Our analysis engages with scholarship on agency in datafied societies \citep{hepp2024agency} and privacy cynicism \citep{lutz2020data} to understand how users navigate environments designed to render resistance futile. Participants who recognized AV surveillance as more comprehensive than smartphone monitoring nonetheless accepted it -- exemplifying the cynical orientation wherein awareness produces resignation. This has troubling implications for governance approaches premised on transparency: if knowledge about surveillance produces acceptance rather than resistance, disclosure requirements may paradoxically reinforce corporate data extraction.

Our analysis situates AV surveillance within broader everyday datafication patterns. Drawing on scholarship examining the `leaky home' \citep{soilen2025leaky}, `invisible witnesses' in smart environments \citep{urquhart2022policing}, and the surveillance ecology of mobile platforms \citep{sophus2020proxy}, we demonstrate how AV monitoring represents a mobile extension of normalized data extraction. Users who have accepted comprehensive smartphone and smart home surveillance encounter AV datafication as another node in established surveillance infrastructure. \citet{barassi2020datafied}'s analysis illuminates how AV surveillance participates in immediacy, archival, and predictive logics of data capitalism, creating longitudinal profiles that persist indefinitely.

We highlight how geographic restrictions on data access create asymmetries impeding informed user understanding, underscoring the importance of universal data access rights as a precondition for meaningful privacy deliberation. Without ability to examine what data manufacturers collect, users cannot develop accurate mental models or identify potential harms \citep{cheong2022smart}. \citet{hepp2024agency} suggest meaningful agency requires not just access to information but institutional conditions enabling effective action -- implying governance must address both transparency and power asymmetries constraining user response.

Practically, we propose governance interventions: comprehensive transparency requirements, data minimization standards, universal log access, and limits on third-party sharing. Design interventions should make data flows visible through contextual indicators and real-time displays. Together, these approaches resist treating privacy as merely an engineering problem, surfacing political choices embedded in datafied systems. \citet{mcguigan2023after} caution that without interventions into underlying business models, companies will develop technical solutions nominally complying with privacy requirements while preserving extractive practices. Governance must anticipate corporate `cynical resignation' and establish binding requirements that cannot be circumvented.

Design interventions alone cannot address structural incentives driving data extraction. Contemporary machine learning's `brute force' approach \citep{stilgoe2018machine} creates persistent pressure for comprehensive surveillance. Governance must establish binding minimums preventing race-to-the-bottom dynamics rather than relying on voluntary restraint. {\citet{rosso2020chilling} demonstrate surveillance produces measurable effects on user behavior, but our findings suggest these effects in AV contexts take the form of resigned compliance. Breaking this pattern requires governance altering the conditions under which users make privacy decisions.

As AVs transition from novelty to infrastructure, their surveillance capacities will shape mobility for millions. Ensuring these technologies serve public interests requires sustained attention to privacy politics. Our findings suggest the urgency of intervention: user indifference reflects not informed consent but successful platform persuasion --- the corporate cultivation of digital resignation that \citet{draper2019corporate} identify as deliberate strategy for neutralizing privacy concern}. Without governance action, the datafied car will entrench surveillance capitalism deeper into everyday life. The challenge for scholars and policymakers is creating conditions where users can exercise meaningful agency over vehicular data \citep{hepp2024agency} --- conditions requiring not just transparency but transformation of institutional arrangements that currently render privacy protection subjectively futile \citep{lutz2020data}.

\end{document}